\documentclass[aps,prl,twocolumn,groupedaddress]{revtex4}
\usepackage{epsfig}
\newcommand{\be}{\begin{equation}}
\newcommand{\ee}{\end{equation}}
\newcommand{\ba}{\begin{eqnarray}}
\newcommand{\ea}{\end{eqnarray}}
\newcommand{\pr}{{\rm pr}}

\begin{document}

{\bf Comment on ``Inferring Statistical Complexity"}

Nearly 30 years ago, in \cite{Crutch-young1} the authors proposed some supposedly novel measures
of time series complexity, and their relations to existing concepts in nonlinear dynamical systems.
At that time it seemed that the multiple faults of this paper would make it obsolete soon. Since
this has not happened, and these faults still infest the literature on what is now called ``computational 
mechanics" (CM) \cite{Shalizi-crutch}, I want here to rectify the situation.

(i) In \cite{Crutch-young1} a R\'enyi graph complexity $C_\alpha$ was defined, and it was proposed
that it is related to the well known R\'enyi entropies $h_\alpha$. Unfortunately, no such connection
exists for general $\alpha$, since the R\'enyi index $\alpha$ has completely different meanings in 
both. Indeed, $C_\alpha$ with $\alpha \neq 0,1$ has not fond any application so far. 

(ii) Both the complexity measure $C_1$ (called ``statistical complexity" in \cite{Crutch-young1})
and ``$\epsilon$-machines" had been introduced previously in \cite{Grass86}. $C_1$ is just what was
called ``forecasting complexity" (FC) in 
\cite{Grass86a,Grass87,Grass88,Zambella}, and ``$\epsilon$-machines" had been called 
minimal deterministic automata. In \cite{Grass86} it had also been proven that FC 
is bounded from below by what is called ``excess entropy" {\bf E} in \cite{Crutch-young1} (the mutual 
information between past and future \cite{Grass86}), and that this bound is in general not saturated. 
In contrast, in \cite{Crutch-young1} it was claimed that FC and {\bf E} are ``simply
proportional".

(iii) While it was not pointed out explicitly, the way how the nodes of the ``$\epsilon$-machine"
were constructed (as equivalence classes of elements of a partitioning in trajectory space) implies
that they also are elements of a partitioning -- as stated explicitly, e.g., in \cite{Shalizi-crutch}. 
This is true in simple cases, but not in general. There 
exist very simple models \cite{Zambella,Grass88,Grass2017} where they are elements of a covering
in which trajectories are multiply covered. This is presumably the most serious mistake, as
the claim that these nodes (called ``causal states" in \cite{Shalizi-crutch}) are elements of a 
partitioning is repeated until now in the CM literature, and makes e.g. 
several proofs in \cite{Shalizi-crutch} obsolete.

\begin{figure}
\begin{center}
\psfig{file=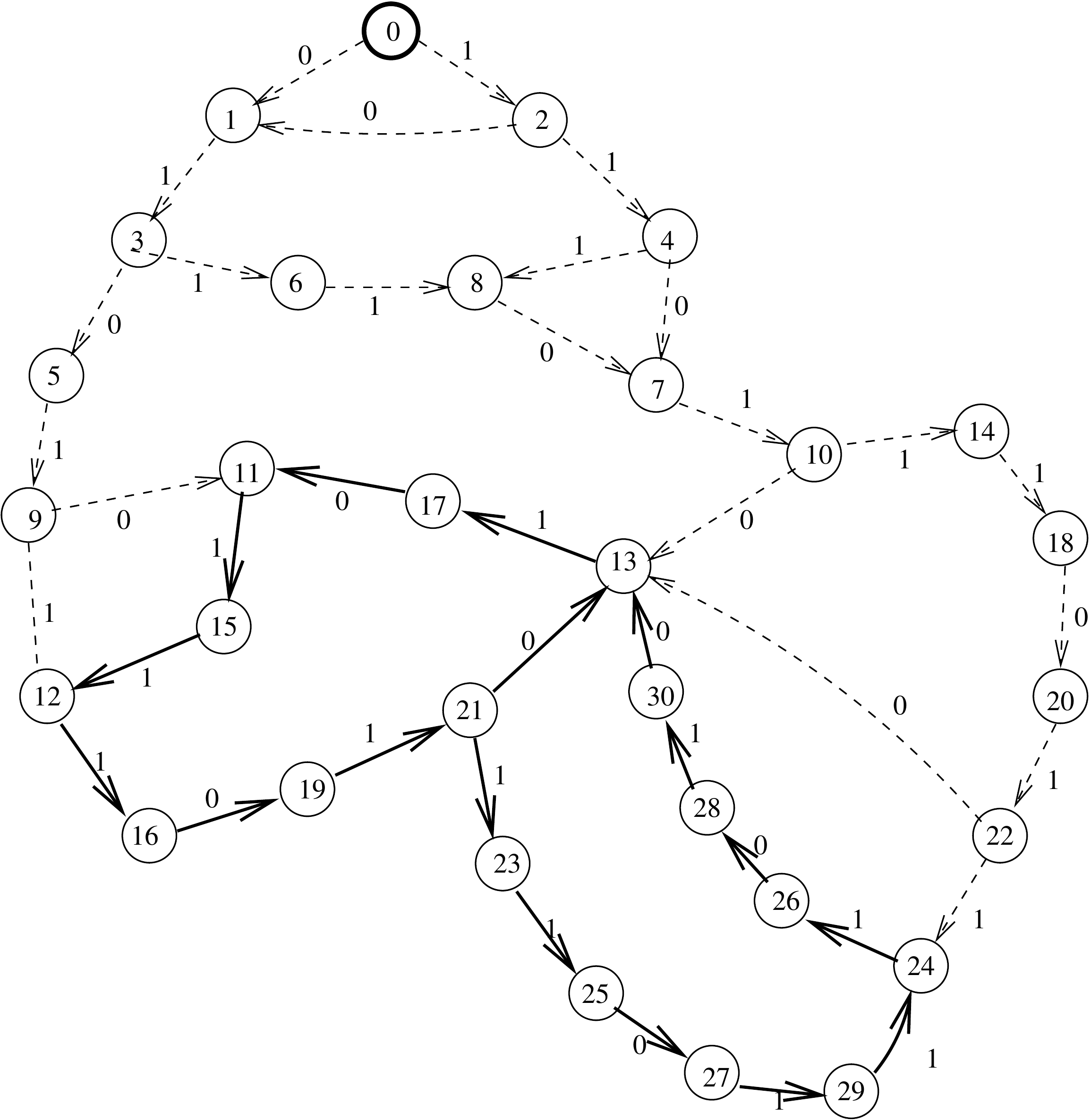,width=7.5cm, angle=0}
\caption{Minimal deterministic (unifilar) graph (``$\epsilon$-machine") which accepts/produces all 0/1 
   sequences on the Feigenbaum attractor up to length 16. Solid lines correspond to the recurrent part, 
   dashed ones to the transient. Notice that this is also the minimal graph which predicts
   their probabilities correctly, since each non-trivial branching ratio is 1:1 except the one at start,
   where $\pr\{0\}:\pr\{1\} = 1:2$. Compare this graph
   (with 15 transient and 16 recurrent nodes) to the graph shown in \cite{Crutch-young1}
   which has 47 recurrent nodes.}
\end{center}
\end{figure}

(iv) Figure 1a in \cite{Crutch-young1} shows supposedly the ``$\epsilon$-machine" that corresponds 
to the length 16 cylinder set of the critical Feigenbaum attractor. Unfortunately, it was not said 
whether it should correspond to the trajectories {\it on} the attractor or in its basin of attraction.
The latter had been given in \cite{Grass88}, while the former is shown in Fig.~1. They are
both different. Indeed, all algorithms for constructing $\epsilon$-machines from
finite data used in the CM literature up to $\sim 2005$ are 
wrong, while the correct algorithm had been given in \cite{Zambella}. In that paper, also an efficient 
algorithm for computing {\bf E} had been given -- the supposed unavailability of such an algorithm was 
considered a major problem in the CM literature until $\sim 2005$.

(v) The marked single-peaked structure of Fig.~2 in \cite{Crutch-young1} results from the fact that 
probabilities and entropies were simply
estimated from fixed length trajectory pieces. A much more careful graph of a similar quantity 
(called ``set complexity" in \cite{Grass86,Grass88}) had been given already in \cite{Grass88}, 
and it displays a much richer and more complex structure.

(vi) In contrast to what its title says, no attempt was made in \cite{Crutch-young1} to actually 
{\it infer} FC -- nor was made such an attempt in any later paper on CM. As discussed in 
\cite{Zambella,Grass2017}, this is not by chance, as inferring FC from imprecise or measured data 
(as opposed to computing it for a precisely given model) is very difficult and so far unsolved.
That is also why only set complexity was estimated in \cite{Grass88}.

More details on CM, ``$\epsilon$-machines", and complexity measures are given in \cite{Grass2017}.

Peter Grassberger \\
JSC, FZ J\"ulich, D-52425 J\"ulich, Germany

\end{document}